# In-Situ Ambient Preparation of Perovskite-Poly(L-Lactide Acid) Phosphors for Highly Stable and Efficient Hybrid Light-Emitting Diodes

Yanyan Duan,[a,b] Guang-Zhong Yin,[a] De-Yi Wang*[a] and Rubén D. Costa*[c]

[a] IMDEA Materials Institute, Calle Eric Kandel 2, 28906 Getafe, Spain

E-mail: deyi.wang@imdea.org

[b] Departamento de Ciencia de Materiales, Universidad Politécnica de Madrid, E.T.S. de Ingenieros de Caminos, Profesor Aranguren s/n, Madrid 28040, Spain

[c] Chair of Biogenic Functional Materials, Technical University of Munich, Schulgasse 22, Straubing D-94315, Germany

E-mail: ruben.costa@tum.de

**KEYWORDS**: metal halide perovskite, bio-polymers, in-situ ambient preparation, perovskite phosphors, hybrid light-emitting diodes

**ABSTRACT**

Metal halide perovskites (MHPs) based phosphor-converted light-emitting diodes (pc-LEDs) are limited by the low MHP stability under storage/operation conditions. A few works have recently stablished the *in-situ* synthesis of MHPs into polymer matrices as an effective strategy to enhance MHP's stability with a low-cost fabrication. However, this is limited within petrochemical-based polymers. Herein, the first *in-situ* ambient preparation of highly luminescent and stable MHPs–bio-polymer filters (MAPbBr$_3$ nanocrystals as emitter and poly(L-lactide acid) (PLLA) as matrix) with arbitrary areas (up to *ca*. 300 cm$^2$) is reported. The MAPbBr$_3$-PLLA phosphors feature a narrow emission (25 nm) with excellent photoluminescence quantum yields (>85%) and stabilities under ambient storage, water, and thermal stress. This is corroborated in green pc-LEDs featuring a low efficiency roll-off, excellent operational stability of ca. 600 h, and high luminous efficiencies of 65 lm W$^{-1}$ that stand out the prior art (e.g., average lifetime of 200 h at 50 lm W$^{-1}$). The filters are further exploited to fabricate white-emitting pc-LEDs with efficiencies of *ca*. 73 lm W$^{-1}$ and x/y CIE color coordinates of 0.33/0.32. Overall, this work stablishes a straightforward (one-pot/*in-situ*) and low-cost preparation (ambient/room temperature) of highly efficient and stable MHP-bio-polymer phosphors for highly performing and more sustainable lighting devices.

# 1. INTRODUCTION

Artificial lighting is responsible for about 20% of the worldwide energy consumption and nearly 7% of the global $CO_2$ generation.[1] Thus, it is mandatory to search for low-cost and efficient lighting sources. In this context, phosphor-converted light-emitting diodes (pc-LEDs) represent a leading example that is slowly ruling the market.[2-6] In pc-LEDs, the blue InGaN chip is combined with the so-called phosphors, leading to primary electroluminescence and secondary photoluminescence named color down-conversion emission, respectively.[7-9] Herein, color down-converting filters or phosphors are recognized as the key components to realize highly efficient pc-LEDs, as the blue InGaN chip has been already optimized.[10]

Metal halide perovskites (MHPs) with a typical structure of $ABX_3$ (where A is $Cs^+$, $CH_3NH_3^+$ (MA), $CH(NH_2)_2^+$ (FA); B = $Pb^{2+}$; X = $Cl^-$, $Br^-$, and $I^-$) are emerging as promising phosphors due to their excellent merits, such as *i*) easy synthesis methods to cover the whole visible range,[11-13] *ii*) high photoluminescence quantum yields (PLQYs),[14-15] and *iii*) narrow full width at half maximum (FWHM) emission bands.[16] However, MHPs based pc-LEDs have been strongly limited by the low stability of the MHPs under ambient and device operation conditions.[10,17] Up to date, several strategies have been developed towards stable MHP-phosphors: *i*) metal oxides/polymers to encapsulate the MHPs,[7,18-23] *ii*) adopting compositional engineering to optimize the Goldschmidt tolerance,[24-25] and *iii*) surface engineering to passivate the surface defects and to increase resistance against, for example, polar solvents.[26-27] However, the most common strategy to fabricate MHP-phosphors is the post-encapsulation of the as-prepared MHPs with polymer matrices like polymethylmethacrylate (PMMA),[12,28-32] or resin.[19-20] For example, Sun and colleagues described the preparation of color converters through dispersing green- and red-emitting MHP@SiO$_2$ powders into a PMMA matrix, meeting high efficiencies and stabilities – *i.e.,* 61.2 lm W$^{-1}$ and 227 h, respectively.[30]

Regardless of the type of MHP emitters, the most serious issue is the homogeneous physical blending of MHPs into the polymer matrix or resin without affecting their excellent photoluminescence features. A few warnings about the fabrication of MHP-polymer phosphors are the use of anhydrous organic solvents to avoid the moisture-induced degradation of the MHPs,[33] and the need to enhance the polymer-MHP interfacial contact to avoid phase separation due to the strong aggregation between MHPs.[34-35] Finally, the synthesis of MHPs is not easily up-scaled, since this requires inert atmospheres and repetitive processes of hot injection, centrifugation, re-dispersion, and vacuum drying. Therefore, it is of utmost relevance to develop facile and more reliable strategies to *in-situ* fabricate stable MHPs-polymers composite films under ambient conditions.

In this context, several groups have focused on the *in-situ* synthesis of MHPs in polymers media.[36-38] The prior-art applied to pc-LEDs consists of a handful number of examples.[39-45] For instance, Zhong's group developed an *in-situ* fabrication strategy to prepare MAPbX$_3$-polyvinylidene fluoride (PVDF) composite, dissolving MABr, PbBr$_2$, and PVDF powder in DMF and controlling its evaporation to separate the crystallization processes of PVDF and MAPbX$_3$.[40] The MAPbBr$_3$-PVDF films feature PLQY of 95 % and stabilities with respect to water (400 h) and UV irradiation stress (8% decrease of PLQY under 365 nm irradiation for 400 h). However, the composite film is highly unstable at temperatures higher than 70 °C. This limits the device stability to a few hours, while device efficiency is remarkable (109 lm W$^{-1}$). Meanwhile, the MAPbBr$_3$-polystyrene (PS), MAPbBr$_3$-polycarbonate, MAPbBr$_3$-polyvinyl chloride (PVC), and MAPbBr$_3$-acrylonitrile butadiene styrene films have also been prepared adopting a swelling-deswelling microencapsulation strategy, through spin-coating the MAPbBr$_3$ precursor on the commercial polymer films.[41] These composite films show limited PLQYs (16-48%), but excellent water stability, revealing less than 7% decay in PLQYs after being immersed in water for 60 days in average. They were applied to pc-LEDs as proof of concept without disclosing device stability or efficiency. Similarly, *quasi*-2D PEA$_2$(MAPbBr$_3$)$_2$PbBr$_4$-PVC composites were fabricated through inkjet printing the MHPs precursors on the polymer substrates.[42] These films show a PLQY of 65%, excellent environmental stability-*e.g*., 70% and 80% of the initial emission intensity remains after being exposed to water and air for 50 days, respectively. The authors did not show device applications. Likewise, stable and flexible CsPbBr$_3$-ethylene vinyl acetate composite films were prepared through an *in-situ* one-step strategy.[39] Finally, MHP-polymers (PMMA, poly(butyl methacrylate), and PS),[43] CsPbBr$_3$@PS fiber film,[44] and CsPbBr$_3$-PMMA composite materials[45] have also been reported using *in-situ* preparation through thermal/UV polymerization, electrospinning, and microfluidic spinning microreactors, respectively.

Besides the above successful *in-situ* preparation and use of MHP-polymer phosphors, the device performance does not meet high efficiency and stability values yet. In addition, all these works have been limited to petrochemical-based polymers, while bio-based and biodegradable polymers have been largely neglected so far. It is here where the major thrust of this manuscript sets in. Poly(L-lactide acid) (PLLA) is proved to be an alternative to petrochemical-based polymers in many applications. It is derived from non-toxic renewable resources, like corn starch, rice, or sugar cane, and is also bio-degradable.[46] Compared with other both bio-based and bio-degradable polymers counterparts, like poly(hydroxyl alkanoate) (PHA), PLLA shows better processibility, can be processed by injection molding, film extrusion, blow molding, thermoforming, fiber spinning, and film forming.[47] Most of all, solubility parameter predictions indicate that PLA will interact with nitrogen compounds (like $CH_3NH_3$) and will not interact with water.[47]

Herein, we report on the *in-situ* ambient fabrication of MHPs–bio-polymer phosphors based on $MAPbBr_3$ nanocrystal (NCs) as emitter with PLLA as matrix. This strategy is designed through transferring the $MA^+$, $Pb^{2+}$, $Br^-$ from soluble status (in N, N-dimethylformamide, DMF) to supersaturated status (in dichloromethane, DCM) without the help of inert gas. The as-prepared self-standing composite films exhibit a homogenous morphology combining the best of both components, that is, excellent emission features of $MAPbBr_3$ NCs (narrow emission of 25 nm and PLQYs of >85%) and high transmittance of PLLA matrices (ca.85% over the visible range). In addition, PLLA also provides outstanding resistance against water, ambient stresses compared to those of the prior art – *i.e.*, 65% and 85% of the initial emission intensity has been kept after being immersed in water and exposed to air for 1 month, respectively. This is further combined with an excellent thermal stability, remaining >80% of the initial emission without any change in the emission band shape after 24 h at 60 °C under ambient conditions. These outstanding figures allowed us to fabricate highly stable and efficient green and white pc-LEDs. On one hand, green pc-LEDs outperform the above discussed state-of-the-art (see Table S1) with operational stabilities of ca. 600 h and high luminous efficiencies of 65 lm W$^{-1}$. On the other hand, this is also reflected in pc-WLEDs combining MAPbBr$_3$-PLLA composite films and commercial red emissive K$_2$SiF$_6$:Mn$^{4+}$ (KSF) phosphor. These devices exhibit efficiencies of 72 lm W$^{-1}$ with a x/y CIE color coordinates of 0.33/0.32 and a color correlated temperature of 5410 K. Overall, this work builds up on the current art stablishing a straightforward (one-pot/*in-situ*) and low-cost preparation (ambient/room temperature) of highly efficient and stable MHP-bio-polymer phosphors for highly performing and more sustainable lighting devices.

**2. EXPERIMENTAL SECTION**

**2.1. Chemicals.** Methylammonium bromide (MABr, > 99.99%) was purchased from Greatcell Solar Materials. Lead bromide (PbBr$_2$, trace metal basis, 99.999%), N, N-Dimethylformamide (DMF, anhydrous, 99.8%), Oleic acid (OA, technical grade, 90%), Oleylamine (OAm, technical grade, 90%) were obtained from Sigma-Aldrich. Dichloromethane (DCM, HPLC, > 99.99%) was supplied by TCI. PLLA (2003D, Mn

= 155,500, MFI = 2.9 g 10 min$^{-1}$) was purchased from Nature Works. $K_2SiF_6:Mn_4^+$ (KSF) powders were purchased from Grirem Advanced Materials Co.,Ltd, China.

**2.2. Synthesis.** PLLA (1 g) was dissolved into DCM (10 mL) in the glass vial upon stirring for 3 h. This leads to a homogenous and transparent solution (PLLA-DCM solution). In parallel, MABr (0.16 mmol) and PbBr$_2$ (0.16 mmol) were dissolved in DMF (5 mL). In addition, OA (0.5 mL) and OAm (25 µL) were used to stabilize the precursor solution, and this precursor solution was kept stirring for 1 h. Then, different volumes (50 µL, 100 µL, 150 µL, 200 µL, and 250 µL) of the above precursors were added dropwise into the PLLA-DCM solution (MAPbBr$_3$-PLLA-DCM solution). A strong green emission is immediately observed. Here, the MAPbBr$_3$ forms through the ligand-assisted reprecipitation (LARP) process.[28] Under this condition, DMF is selected as the good solvent to dissolve the MABr and PbBr$_2$, while DCM as the bad solvent (anti-solvent) to initiate the formation of MAPbBr$_3$. It is important to highlight the unusual reaction conditions at room temperature and under ambient environment without any inert gas protection. During the above process, the DCM solvent was transferred by the measuring cylinder (20 mL); OA, OAm, and the perovskite precursor solution were added by the Gilson pipettes.

After continuous stirring for 2 h, the MAPbBr$_3$-PLLA-DCM solutions were deposited onto a glass petri dish and slowly dried in a fume hood overnight. Here, the obtained films are named as 50 µL MAPbBr$_3$-PLLA, 100 µL MAPbBr$_3$-PLLA, 150 µL MAPbBr$_3$-PLLA, 200 µL MAPbBr$_3$-PLLA, 250 µL MAPbBr$_3$-PLLA, respectively. These films all show good transparency as expected from PLLA, and a good dispersion of the perovskite nanoparticles. Herein, the thickness of the films can be tuned by the amount of solution/concentration and the size of the petri dishes used to dry the solution under ambient conditions inside the fume hood. The films with a thickness of ca. 0.5 mm (obtained from 10 mL solution dried in the petri dishes with a diameter of 5 cm) are selected for further tests based on their excellent optical properties.

**2.3. Fabrication of pc-LEDs.** The MAPbBr$_3$-PLLA based pc-LEDs were fabricated through putting the films on top of the commercial blue-emitting LED chip (3 W, 450 nm) in a remote configuration (1.5 cm between the LED chip and the films). pc-WLEDs were fabricated as follows. KSF (100 mg) was encapsulated into the UV cure adhesion (Ossila, E132) as the red emissive emitter. The mixture was coated on a piece of quartz glass (2.5 cm*2.5 cm), followed by a UV-curing process at room temperature for about 1 min. Finally, the pc-WLEDs were obtained by putting the above KSF-adhesion layer and the MAPbBr$_3$-PLLA layer on a commercial blue-chip LED (3 W; 450 nm) in sequence. Herein, the 150 μL MAPbBr$_3$-PLLA films with a thickness of ca. 0.5 mm are chosen as the green filter. The distance between the LED chip and the filters is set to 0.5 cm.

**2.4. Characterizations.** The UV-Vis absorption spectra were performed on a UV-2600 spectrometer (SHIMADZU). X-ray powder diffraction (XRD) measurements were obtained using the Philip X' Pert PRO diffractometer, using Ni filter and Cu Kα radiation source ($\lambda$ = 0.154 nm). The samples were scanned from 5° < 2θ < 60° with an increment of 2°/min. SEM (scanning electron microscopy) were carried out using the FEI 600i, OXFORD INSTRUMENTS. Transmission electron microscopy (TEM) and High-resolution TEM (HRTEM) were taken from the FEI Talos F200X microscope.

The emission spectra were measured with FS5 Spectrofluorometer (Edinburgh Instruments) using the SC-10 module. The time-resolved PL decay curves were obtained using a time-correlated single photo counting (TCSPC) system with a picosecond pulsed diode laser (EPL). The PLQYs were determined by a SC-30 integrated sphere.

The pc-LEDs were measured using a Keithley 2231A-30-3 as the power supply, and the Avaspec-2048L spectrometer coupled with a calibrated Avasphere 30-Irrad integrated sphere (Avantes) to monitor the change in the electroluminescence spectra, the luminous efficiency, and the emission intensity over time. During the thermal stability measurement, the temperature of the films was preciously monitored by a thermographic camera ETS320 (FLIR systems, Inc).

## 3. RESULTS AND DISCUSSION

**3.1. *In-situ* Ambient Preparation of MAPbBr$_3$-PLLA Phosphors**. We followed the ligand-assisted reprecipitation process for the *in-situ* ambient preparation of hybrid MAPbBr$_3$-PLLA phosphors.[28, 48-49] As illustrated in Figure 1a, the preparation of hybrid MAPbBr$_3$-PLLA composite films consists of three steps. Firstly, a saturated solution of PLLA in DCM (PLLA-DCM solution, process I) was prepared. Non-polar or low polar solvent, like toluene and hexane, are widely used as the anti-solvent for the formation of MHPs.[28, 50] However, PLLA cannot be dissolved in both solvents. Thus, DCM was selected not only as the good solvent for PLLA, but also as a poor solvent for the MAPbBr$_3$ NCs. In parallel, MABr and PbBr$_2$ were dissolved in DMF with the assistance of the Oleic acid (OA) and Oleylamine (OAm) as the stabilizers. Next, different amounts (50, 100, 150, 200, and 250 μL) of MAPbBr$_3$ precursor solutions were added dropwise to the PLLA-DCM solution (process II, MAPbBr$_3$-PLLA-DCM solution forms). During this process, the PLLA matrices provide the confined space for the MAPbBr$_3$ to avoid the aggregation. An intense green emission is observed immediately upon the addition of the MAPbBr$_3$ precursor under UV-irradiation (Figure 1b). Therefore, MAPbBr$_3$ NCs are formed through a rapid recrystallization process. As a reference, the MAPbBr$_3$ precursor was also directly introduced into the DCM solution (MAPbBr$_3$-DCM solution) without the PLLA. As shown in Figure S1, the MAPbBr$_3$-PLLA-DCM solution is very stable, no precipitates are observed even put in the air for long time under ambient condition. However, significant precipitation can be observed when the MAPbBr$_3$-DCM solution is put in air for only 1 h. Finally, the self-standing MAPbBr$_3$-PLLA composite films are obtained after stirring the mixture for another 3 h followed by drying in the fume hood overnight (process III). In this stage, the PLLA layer will provide efficient barrier against moisture and oxygen from the surrounding environment. As shown in Figure 1c, the films with a diameter of 1.6 cm and a thickness of *ca*. 0.4 mm exhibit highly homogenous appearance in terms of color and emission – *vide infra,* highlighting the outstanding compatibility between PLLA and MAPbBr$_3$. In addition, the pure PLLA films exhibit high transmittance (ca.85% over the visible range, Figure S2), serving as an excellent matrix for MAPbBr$_3$ QDs. Finally, Figure 1d shows an easy large-scale preparation of MAPbBr$_3$-PLLA composite films with a size of 8.5 cm × 30 cm. All-in-all, this one-pot/*in-situ*, low-cost, easy-processed ambient preparation technique provides a great opportunity to get easy access to perovskite-phosphors, if highly efficient and stable photoluminescence under stress conditions (ambient storage and water as well as irradiation and temperature used in pc-LEDs) are met.

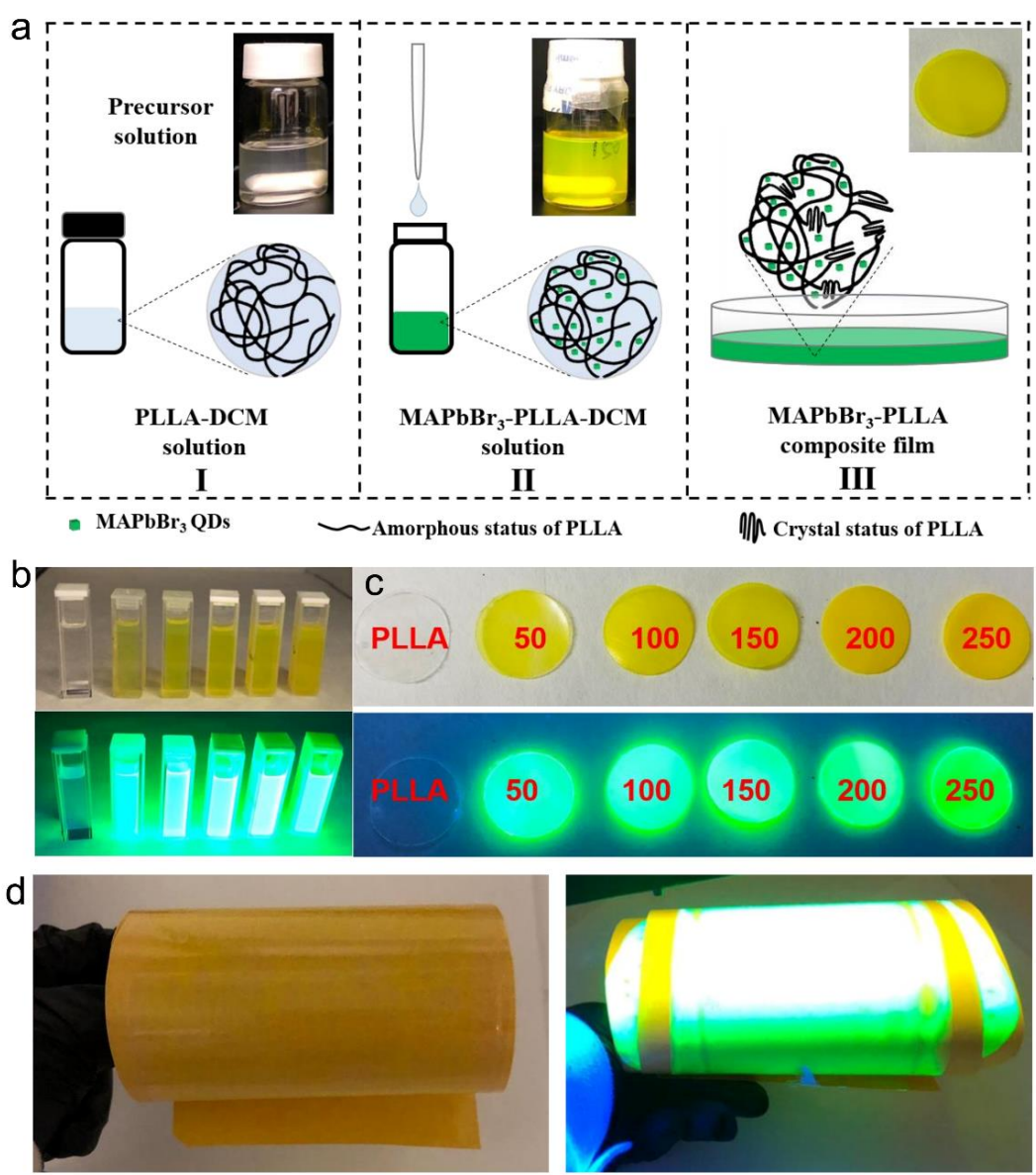

**Figure 1.** (a) Schematic illustration of the *in-situ* ambient preparation of hybrid MAPbBr$_3$-PLLA phosphors; (b) photographs of the MAPbBr$_3$-PLLA-DCM solutions

and (c) photographs of the MAPbBr$_3$-PLLA composite films under ambient light (top) and UV-irradiation (302 nm, bottom); (d) photographs of large-scale prepared MAPbBr$_3$-PLLA composite film with a size of 8.5 cm × 30 cm under ambient daylight (left) and UV-light (302 nm, right).

**3.2. Characterization of MAPbBr$_3$-PLLA Phosphors**. The morphology and composition of the MAPbBr$_3$-PLLA composite films were studied using X-ray diffraction (XRD), X-ray photoelectron spectroscopy (XPS), α-step profilometry, transmission electron microscopy (TEM), and scanning electronc microscopy (SEM). The XRD patterns of the pure PLLA film and the MAPbBr$_3$-PLLA composite films are shown in Figure 2a. The peaks located at 15.18°, 17.08°, 19.51°, 22.78° refer to the PLLA.[51-52] While no obvious MAPbBr$_3$ peaks are detected in 50 μL MAPbBr$_3$-PLLA composite films due to their low content (<0.1 wt%), new diffraction peaks appear at 30.4°, 34.1°, and 43.5° for 150 μL MAPbBr$_3$-PLLA films. They correspond to (200), (210), and (220) planes of the cubic MAPbBr$_3$, respectively.[40, 53] Meanwhile, the intensity of these peaks increases upon using higher volume of the precursors. Nevertheless, the intensity of the diffraction peaks is very weak due to the low content of the MAPbBr$_3$ in the PLLA matrix as well as the coverage effect of the PLLA matrix. XPS spectra further confirm the existence of Br and Pb in the composite film, suggesting the formation of the MAPbBr$_3$ (Figure S3). Further corroboration is provided with TEM images (Figures 2b and 2c). For instance, 150 μL MAPbBr$_3$-PLLA films exhibit a homogenous distribution of the MAPbBr$_3$ particles with an average size of ~ 3.5 nm. High resolution TEM (HRTEM) images (Figure 2d) confirm the crystallinity of the MAPbBr$_3$ nanoparticles featuring a lattice distance of 2.92 nm that is consistent with the (200) plane of the cubic MAPbBr$_3$ (space group: Pm-3m No. 221).[54] In addition, the particle size increases upon using higher volume of precursor (Figure S4). Finally, SEM inspection of the surface of the pure PLLA and MAPbBr$_3$-PLLA composite films were carried out (Figure S5). Pure PLLA films show the typical corrugated surface (Figures S5a and 5b), while the surface is smoothed upon forming MAPbBr$_3$ particles into the polymer network (Figures S5c, S5d and Figure S6). The smoothest surface morphology is obtained with the 150 µL MAPbBr$_3$-PLLA composite films (Figures S5e, S5f and Figure S6), while pin-holes are observed upon increasing the volume of the precursor (Figures S5g and S5h). Importantly, no obvious particles are noted at the surface, while the SEM-EDS mapping images show that both Pb and Br are all well dispersed in the film (Figure S7). Therefore, all these observations demonstrate that the MAPbBr$_3$ nanoparticles are homogenously embedded in the PLLA network through the *in-situ* ambient preparation strategy.

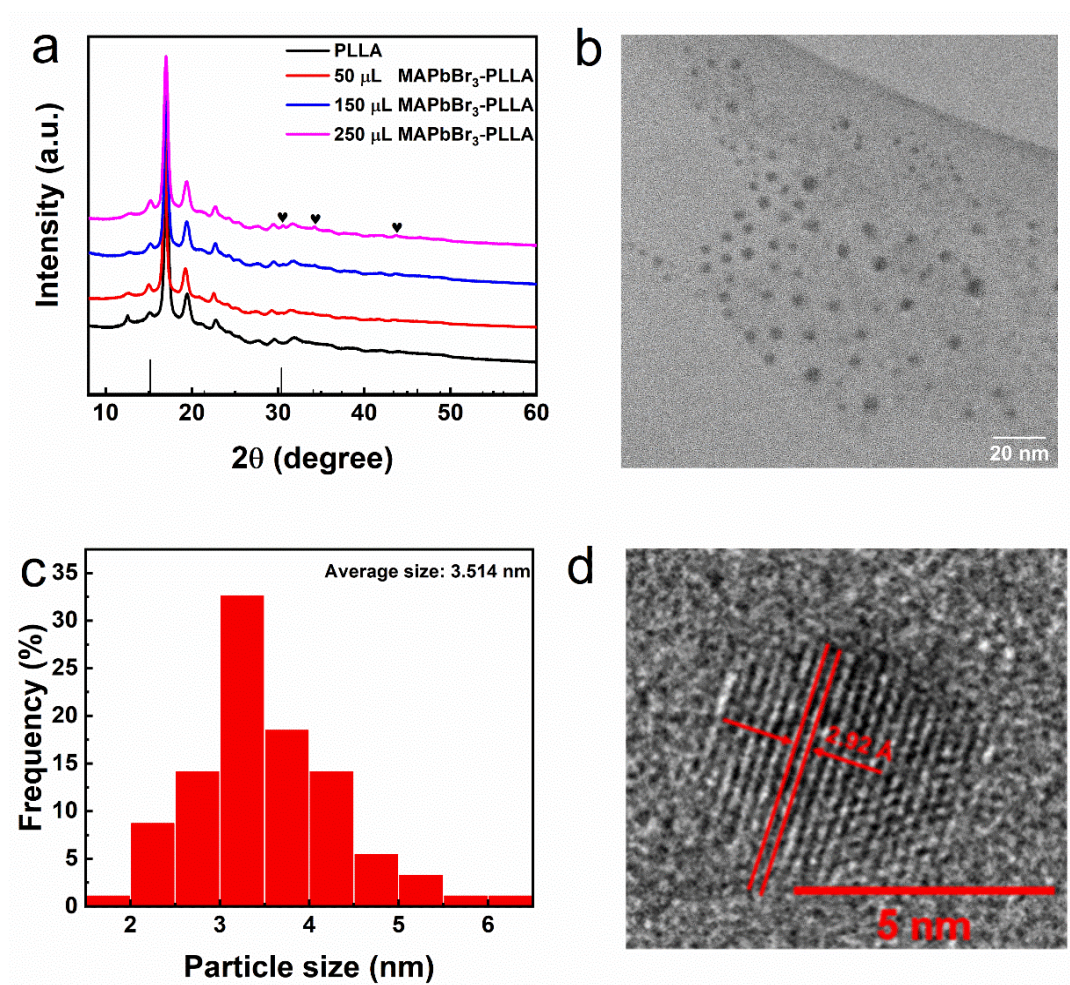

**Figure 2**. (a) XRD spectra of the pure PLLA film and the MAPbBr$_3$-PLLA composite films; ♥ indicates the peaks belong to MAPbBr$_3$; (b) typical TEM image of 150 μL-PLLA composite film, the sample was prepared through dropping the diluted MAPbBr$_3$-PLLA-DCM solution on the grid and dried in the air; (c) histogram analysis of the particle size distribution for the view in Figure 2b; (d) HRTEM image of a typical MAPbBr$_3$ NCs.

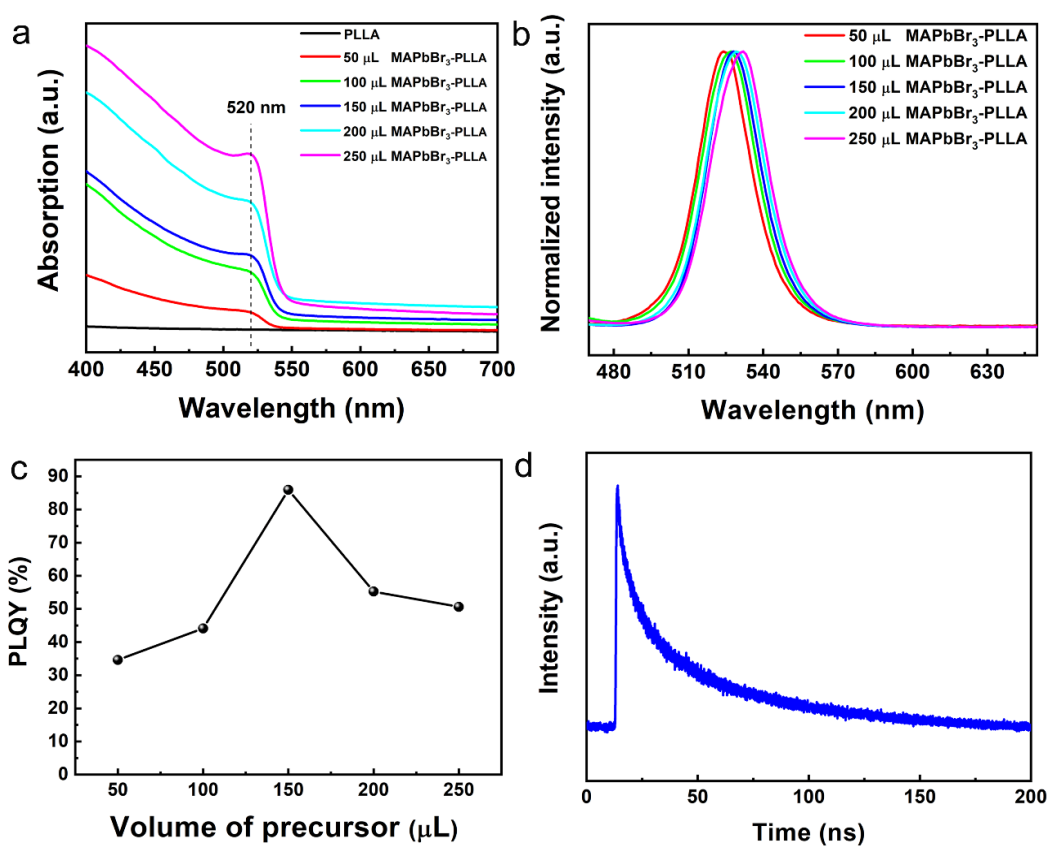

**Figure 3.** Optical properties of PLLA and MAPbBr$_3$-PLLA composite films (see legend): (a) UV-Vis absorption spectra; (b) normalized emission spectra; (c) PLQYs upon increasing volume of the precursor; (d) time-resolved emission decay of the 150 µL MAPbBr$_3$-PLLA composite film.

The optical features of PLLA and MAPbBr$_3$-PLLA composite films were studied using UV-Vis absorption as well as steady-state and time-resolved emission spectroscopy (Figure 3). While pure PLLA films show a featureless absorption and emission spectra, the typical absorption and emission features of the MAPbBr$_3$ nanoparticles are noted for the MAPbBr$_3$-PLLA films.[55] As expected, the absorption band is centered at 520 nm and its intensity increases with the amount of volume precursor. In line with the morphology characterization, there is a perfect match of the UV-Vis spectra taken from 5 different positions in large-area 150 µL MAPbBr$_3$-PLLA composite films (Figure S8), corroborating their homogenous distribution. Concerning the emission features, a sharp green emission band with a FWHM value of ca. 25 nm is generally noted. However, the emission peak shows a gradually red-shift as the increasing volume precursor concentration (Figure 3b and Table S2). This is related to the increase of the crystalline size as noted in XRD and TEM assays – *vide supra*.[36, 41] Thus, we focus on the changes of PLQY values (Figure 3c) to determine the best film composition for lighting applications. We found that the key technical aspect that decides the intensity of the emission is the amount of the introduced precursor. Here, the 150 µL MAPbBr$_3$-PLLA composite film shows the highest PLQY value of >85 %, while still a remarkable PLQY value (>50 %) is measured for the 250 µL MAPbBr$_3$-PLLA composite film (Figure 3c and Table S2). As explained in the previous literature,[56] the increased emission or PLQY at low concentration of the precursor (from 50 µL-PLLA to 150 µL-PLLA) is attributed to the increase in the fluorescence brightness due to a better balanced of surface defect protection and nanoparticle size as stated in the TEM and XRD assays – *vide supra*. However, the decreased emission or PLQY at higher concentration of precursor (from 150 µL-PLLA to 250 µL-PLLA) is attributed to an increase of the scattering and self-absorption events. Finally, time-resolved emission measurements were performed to determine average excite-state lifetimes of 30-60 ns for the composite films (Figure 3d, Figure S9 and Table S2). This is in line with those values of MAPbBr$_3$-based films or nanoparticles.[57-58] The 150 µL

MAPbBr$_3$-PLLA film shows the longest lifetime of 58.03 ns among these films, suggesting the least surface defects that act as the non-reactive recombination sites.[59]

In light of the excellent optical and morphological features of 150 μL MAPbBr$_3$-PLLA films, we proceed to test their stabilities upon storage under ambient conditions, immersed into the aqueous solution, and under thermal stress – *i.e.*, increasing heating temperature and constant heating. As shown in Figure 4a, they show outstanding storage stability (ca. 10% emission loss without emission band changes after one month storage under ambient conditions (Figure S10 and Table S3). Indeed, the composite retains intense green emission after about half year in air (Figure S11). In water (Figure 4b), the emission intensity slowly decreases up to ca. 65% during the first 3 days, holding constant for 1 month. No emission band changes are noted (Figure S12 and Table S3). Thus, these results confirm that bio-based polymer matrices, in general, and PLLA, in particular, are of high interest, since they provide an effective protection against both water and oxygen compared to traditional polymers – see Table S4 for a direct comparison with the literature.[39,41-45]

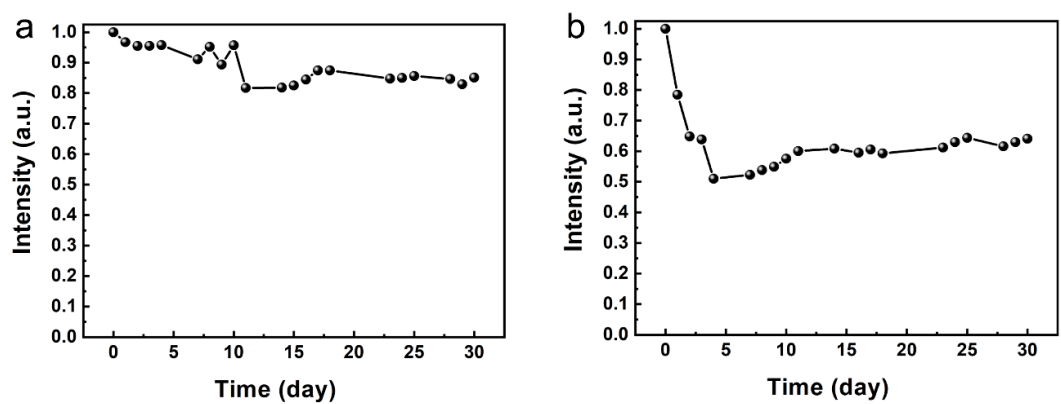

**Figure 4.** Changes in the emission intensity of the 150 μL MAPbBr$_3$-PLLA films over time under the ambient environment (a) and in aqueous solutions (b).

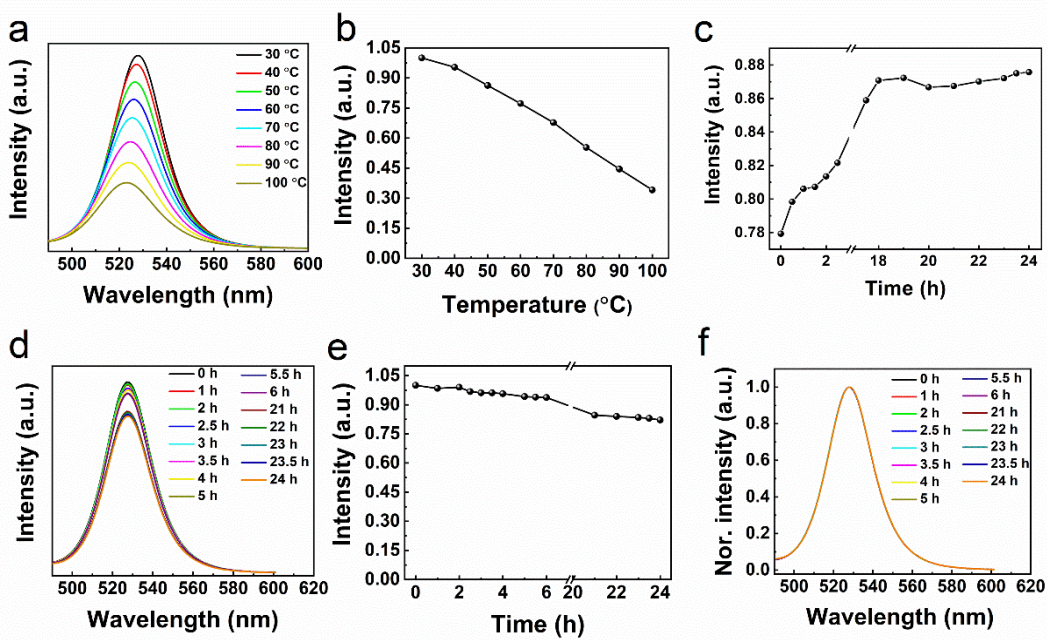

**Figure 5.** Thermal stability of 150 μL MAPbBr$_3$-PLLA films in ambient (air; 60% moisture): (a) and (b) temperature-dependent emission spectra (see legend); (c) emission intensity recovery over time at room temperature; (d) and (e) emission changes over time at 60 °C; (f) normalized emission spectra overtime at 60 °C.

Next, the thermal stability of 150 μL MAPbBr$_3$-PLLA films was firstly tested through heating on a hot plate up to 100 °C in 10 ºC steps, and keeping for 5 min at each measuring temperature. As shown in Figure 5a and 5b, the emission intensity decreases upon increasing temperature, while the emission band is slightly broadened and blue-shifted (Figure S13).[60] Remarkable, the films retain *ca*. 35 % of the initial emission at 100 °C for 5 min. This is in perfect agreement with previous literature, in which the MAPbBr$_3$ nanoparticles were embedded into, for example, PS and polycarbonate matrices.[41] In short, the intensity loss is related to the thermal-induced non-radiative recombination processes. The thermal-induced width broadening is attributed to both exciton-acoustic phonon and exciton-optical phonon.[28, 61] The blue shifted emission relates to the thermal-induced lattice expansion and the renormalization of electron–phonon during the thermal treatment as reported in other MHPs based materials.[28, 41, 61-62] Indeed, both the emission intensity and band shape almost fully recover upon cooling at room temperature (Figure 5c and Figure S13). The irreversible emission loss (~10%) is ascribed to rearrangements at the MAPbBr$_3$/polymer interface.[62-63] Finally, we measured the thermal stability at 60 °C in air. Here, emission band shape holds constant and the emission intensity slightly reduces (<15%) after 24 h (Figures 5d, 5e, and 5f). However, the emission is quickly fully recovered upon cooling at room temperature (Figure S14). Taking into account the prior art in the thermal-stability of MHP-polymers provided in Table S4, the use of bio-based and bio-degradable PLLA matrix stands out to simultaneously prepared and stabilized MAPbBr$_3$ nanoparticles as phosphors for lighting purposes.

**3.3. pc-LEDs with MAPbBr$_3$-PLLA Phosphors.** The pc-LEDs were fabricated using a remote configuration with PLLA and MAPbBr$_3$-PLLA filters and commercial blue-emitting LED chip (3 W, 450 nm) – see Experimental Section for details. The changes of the electroluminescence spectra and the luminous efficiency are monitored upon increasing the current from 1 to 150 mA (Figure 6a and Figure S15). For reference purpose, the characterization of bare commercial LED is provided. The same narrow blue emission located at *ca*. 450 nm and luminous efficiency (ca. 30 lm W$^{-1}$; Figure 6b) are noted for both the commercial and the PLLA film-based LEDs upon increasing the applied current (Figure S15). This is attributed to the excellent transmittance of the PLLA matrix – *vide supra*. The MAPbBr$_3$-PLLA based pc-LEDs show new narrow emission band centered at 528 nm that corresponds to the emission of the MAPbBr$_3$ NCs (Figure 6a). The color down-conversion efficiency holds constant regardless of the applied current, while the luminous efficiency reaches values of ca. 65 lm W$^{-1}$ at 20 mA (Figure 6b). Interestingly, the luminous efficiency reduces to values of 53 lm W$^{-1}$ at 150 mA, highlighting a low efficiency roll-off since the phosphor temperature slightly increases. The operational stability of the MAPbBr$_3$-PLLA based pc-LEDs is assessed recording emission intensity and band shape over time until the 50% of the initial emission intensity (t$_{1/2}$) is reached at 5 mA. As shown in Figure 6c, the emission increases quickly during the first 10 h due to the photo-activation curing of surface trap states.[7, 58] This is followed by a slow exponential decrease over almost 5 days, until reaching a plateau at ca. 60% of the initial intensity for several weeks. The t$_{1/2}$ value is touched at ca. 600 h, representing one of the most stable green pc-LED reported so far – see Table S1 for a direct comparison.[8,18,20,58,64-65]

The slow degradation of the MAPbBr$_3$-PLLA filter is unfortunately irreversible. The photoaged filter shows a blue-shifted emission band (*ca*. 5 nm; Figure S16 and Table S3) associated to a strongly reduced PLQY (50%) and excited state lifetimes (32.03 ns). During the exponential decay (<200 h of device operation), the emission band shape does not change, indicating that the deactivation process is ascribed to both, the photo-assisted ionization[66-67] and the photo-induced desorption of the surface ligands.[8,68] In contrast, the emission band blue-shifts in the plateau stage (200-600 h), suggesting an irreversible photo-etching of the MAPbBr$_3$ nanoparticles.[58] Indeed, the fluctuations in the emission intensity point out the presence of moisture induced hydration processes that initially cures the surface trap states before fully passivating the nanoparticles.[7]

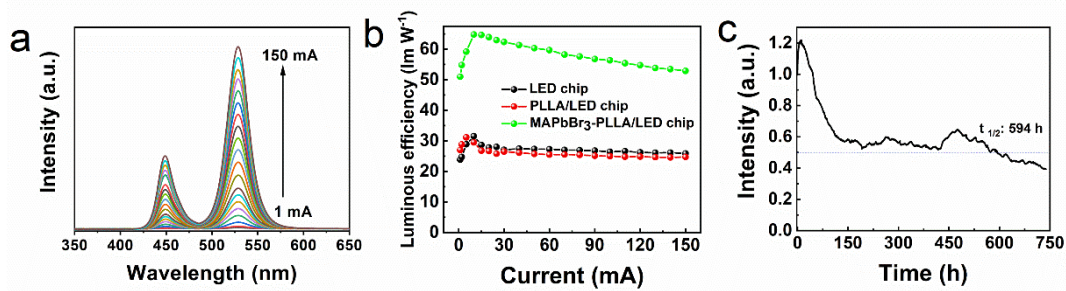

**Figure 6.** Characterization of green pc-LEDs with MAPbBr$_3$-PLLA phosphors: (a) changes of the electroluminescence spectra upon increasing the applied current; (b) luminous efficiencies of the bare LED chip and pc-LEDs with PLLA and the MAPbBr$_3$-PLLA films upon increasing the applied current; (c) the operational stability under ambient conditions at applied current of 5 mA.

Finally, pc-WLEDs were fabricated combining the commercial red KSF phosphor (encapsulated in UV cure adhesive), the MAPbBr$_3$-PLLA phosphors, and the commercial blue LED chip (see Experimental Section and Figure 7a). The electroluminescence spectra of the optimized pc-WLEDs corresponds to the emission of the three components in an intensity ratio for standard white emission – *i.e.,* x/y CIE color coordinates of 0.33/0.32, color correlated temperature of 5410 K (Figures 7b and 7c). Besides, further proper and systematical optimizations should be operated to the Color Rendering Index (CRI, 45) and Duv (0.0101) of the pc-WLEDs, compared with the previous work.[69] Indeed, the color gamut of the device is covering 121.9% of the National Television System Committee (NTSC) standard with a matching ratio of 99.3% (Figure 7c). This holds for the lifetime of the device (Figure 7d). Finally, the excellent white quality is associated to a maximum luminous efficiency of 72 lm W$^{-1}$ at 20 mA.

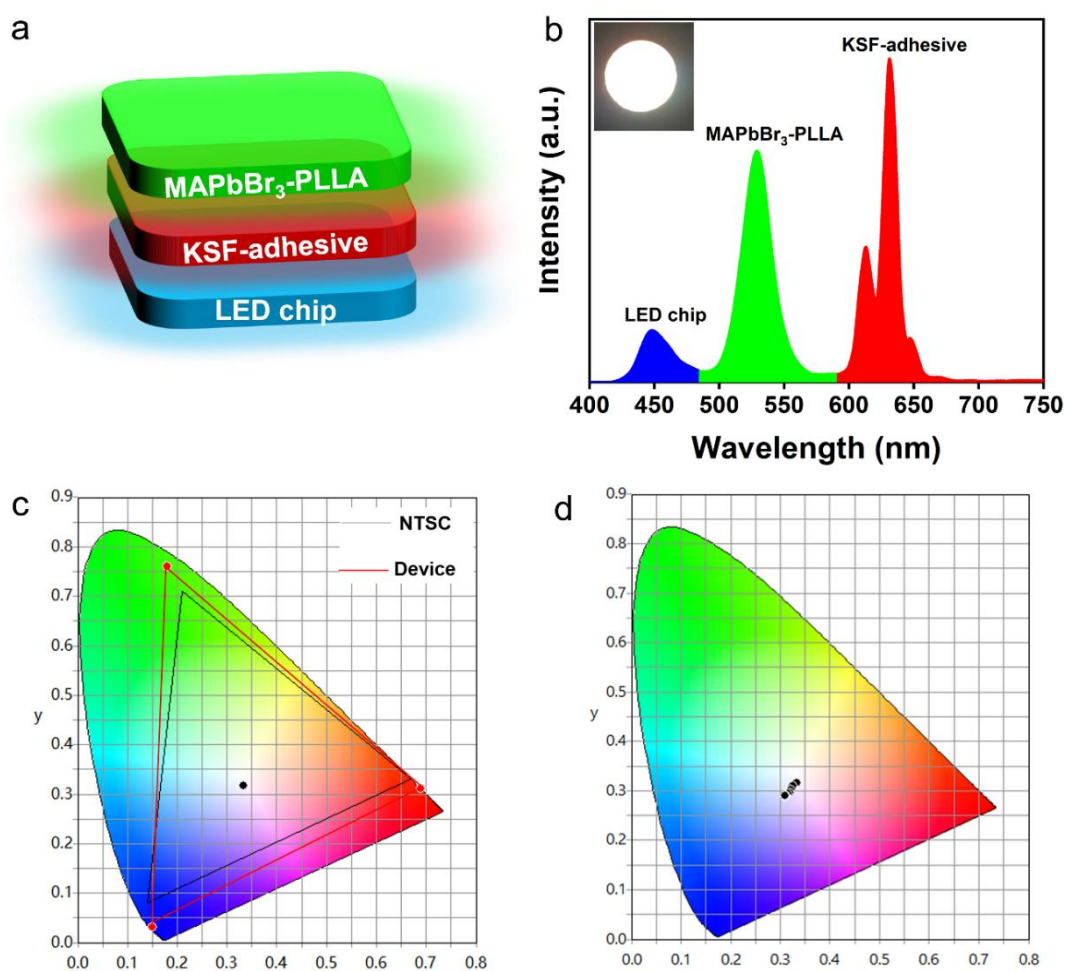

**Figure 7.** Characterization of pc-WLEDs with MAPbBr$_3$-PLLA and KSF phosphors: (a) Scheme of the configuration of the pc-WLEDs device; (b) emission spectra of the pc-WLEDs using green MAPbBr$_3$–PLLA composite film and red KSF-adhesive film; inset is the photographs of a working pc-WLED under 20 mA; (c) x/y CIE color diagram 1931 of the pc-WLED (black point), the color triangle of blue LED chip, green MAPbBr$_3$-PLLA, KSF-adhesive compared to that of the NTSC standard; (d) changes of x/y CIE color coordinate of the pc-WLEDs operating at 20 mA over time.

## 4. CONCLUSIONS

The first stable MAPbBr$_3$-PLLA films were successfully prepared through an on-pot/*in-situ* strategy under ambient/room temperature conditions and applied as phosphors, realizing highly stable and efficient green- and white-emitting pc-LEDs. We demonstrate that this strategy is a straightforward, low-cost, and robust procedure to synthetize highly emissive MAPbBr$_3$ nanoparticles (*ca*. 3 nm; PLQY>85 %) that are homogenously embedded into the PLLA networks with arbitrary areas (up to *ca*. 300 cm$^2$) and effectively shielded, showing similar stabilities against water, storage, and thermal stress to those described in the prior-art. And intense green emission can still be observed after being to the ambient condition for half year. This results in green MAPbBr$_3$-PLLA based pc-LEDs outperforming the current state-of-the-art with luminous efficiencies of *ca*. 65 lm W$^{-1}$ and stabilities of over 600 h. This is further exploited to fabricate pc-WLEDs combining MAPbBr$_3$-PLLA and KSF phosphors. They also feature maximum efficiencies of around 72 lm W$^{-1}$ and excellent wide color gamut covering *ca*. 122% of the NTSC color standard. In light of these results, it is safe to state that the great success in the *in-situ* ambient preparation of the highly emissive and stable MHPs using bio-polymers paves the way towards a more sustainable and highly performing lighting source.

**ASSOCIATED CONTENT**

**SI Supporting Information**

The comparison of the stability of MAPbBr$_3$-PLLA-DCM and MAPbBr$_3$-DCM solutions; the transmittance spectra of the pure PLLA and MAPbBr$_3$-PLLA films; more HRTEM images, UV-Vis absorption spectra, time-resolved emission decay curves, emission spectra; the XPS spectra, the α-step roughness results, the SEM images and the SEM-EDS mapping of films; the photographs of fresh and 172 days old 150 μL MAPbBr$_3$-PLLA film; the temperature-dependent maximum emission wavelength and FWHM. Tables about the emission properties of the prepared films, and direct comparison of the operational stability of the green pc-LEDs as well as the *in-situ* fabricated MHPs-polymers.

**AUTHOR INFORMATION**

**Corresponding Authors**

De-Yi Wang - IMDEA Materials Institute, Calle Eric Kandel 2, 28906 Getafe, Spain;

E-mail: deyi.wang@imdea.org    orcid.org/0000-0002-0499-6138

Rubén D. Costa - Chair of Biogenic Functional Materials, Technical University of Munich, Schulgasse 22, Straubing D-94315, Germany

E-mail: ruben.costa@tum.de    orcid.org/0000-0003-3776-9158

**Authors**

Yanyan Duan - IMDEA Materials Institute, Calle Eric Kandel 2, 28906 Getafe, Spain; Departamento de Ciencia de Materiales, Universidad Politécnica de Madrid, E.T.S. de Ingenieros de Caminos, Profesor Aranguren s/n, Madrid 28040, Spain

orcid.org/0000-0003-4988-4196

Guang-Zhong Yin - IMDEA Materials Institute, Calle Eric Kandel 2, 28906 Getafe, Spain   orcid.org/0000-0003-0065-2706

**Notes**

The authors declare no competing financial interest.

## ACKNOWLEDGMENTS

Y. Y. D. thanks the financial support from China Scholarship Council (CSC, No. 201808440326).

**Table of Contents (TOC)/Abstract Graphic**

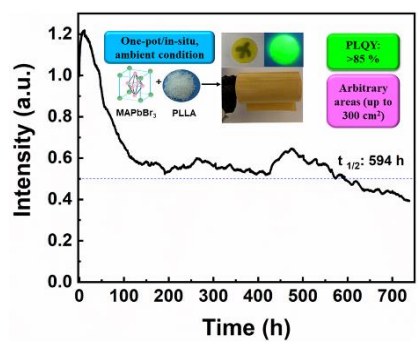